\documentstyle[preprint,aps,eqsecnum,psfig,floats]{revtex}
\def\SH{Stillinger-Helfand}
\def\z{\tilde{z}}

\input epsf.sty

\begin{document}

\title{Cluster Monte Carlo study of multi-component fluids of the
Stillinger-Helfand and Widom-Rowlinson type}

\author{Rongfeng Sun$^\dag$ and Harvey Gould}
\address{Department of Physics, Clark University,
Worcester, MA 01610-1477}

\author{J. Machta}
\address{Department of Physics and
Astronomy, University of Massachusetts, Amherst, MA 01003-3720}

\author{L. W. Chayes}
\address{Department of Mathematics, University of California, Los
Angeles, CA 90095-1555}

\maketitle
\begin{abstract}Phase transitions of fluid mixtures of the type
introduced by Stillinger and Helfand are studied using a continuum
version of the invaded cluster algorithm. Particles of the
same species do not interact, but particles of different
types interact with each other via a repulsive potential. Examples
of interactions include the Gaussian molecule potential and a
repulsive step potential. Accurate values of the critical
density, fugacity and magnetic exponent are found in two and three
dimensions for the two-species model. The effect of varying the
number of species and of introducing quenched impurities is also
investigated. In all the cases studied, mixtures of $q$-species are
found to have properties similar to
$q$-state Potts models.
\end{abstract}
\pacs{05.50.+q,64.60.Fr,75.10.Hk}

\section{Introduction}
Several years ago Stillinger and
Helfand\cite{StHe,StHe2} introduced a simple but nontrivial model
of fluid demixing. Their original model consists
of a binary mixture of
$A$ and $B$ particles. Particles of the same type do not
interact with one another, but $A$ and $B$ particles interact with
a repulsive potential such that the Mayer $f$-function is a
Gaussian. This choice for the $AB$ potential, known as the
Gaussian molecule potential, greatly simplifies the calculation of
virial coefficients and most work for this potential has been done
using series methods\cite{BaMaRo,LaFi}. The main motivation for
this work is to confirm Ising universality for the critical
exponents of continuum systems.

In this paper we study the
\SH\ model and some of its generalizations using cluster
Monte Carlo methods. Where possible, we compare our results to
the series analyses and to results for the Ising-Potts universality
classes. Although the Gaussian molecule potential
yields a more tractable virial expansion, it is easier to
implement a cluster algorithm for the repulsive step potential. We
also consider a generalization of the \SH\ model to $q$ species
(components), such that particles of the same species do not
interact but particles of different species interact with a
repulsive potential. We expect that this generalization will be in
the same universality class as the
$q$-state Potts model for $q$ not too large and another motivation
for this study is to confirm this correspondence. For example, we
know that the two-dimensional (2D) Potts model for $q>4$
has a first-order transition. Does the
$q$-component 2D \SH\ model also have a first-order transition for
$q>4$? In addition, we consider the effect of quenched disorder by
randomly adding fixed scattering centers. There are general
arguments\cite{HuBe} that quenched disorder causes first-order
transitions to become continuous. These arguments hold
rigorously for 2D Potts models\cite{AiWe}, but have not been
studied for continuum models.

In previous work\cite{JGMC} cluster Monte Carlo methods were
applied to the Widom-Rowlinson model\cite{WiRo}. The
Widom-Rowlinson and \SH\ models are closely related, the only
difference is that the Widom-Rowlinson model has a hard-core
interaction between different species. In this paper, the
invaded cluster Monte Carlo method introduced in Ref.\cite{JGMC}
is extended to soft-core repulsive potentials and is used to find
the phase transition point for a given temperature without prior
knowledge of the critical fugacity. The invaded cluster method has
almost no critical slowing for the Widom-Rowlinson model, and we
find that similar results hold for the
\SH\ models studied here.

\section{Description of the Models and Notation}
We consider $q$-component ($q
\geq 1$) fluids in
$d$ dimensions with $d=2,3$. The components (species) have no
self-interaction but particles of one species interact with
particles of all other species via an isotropic repulsive
potential,
$U(r)$. We consider two choices for $U(r)$,
\begin{mathletters}
\label{eqn:allpots}
\begin{equation}
U_{\rm step}(r) = \cases{U_0, & ${\rm if\ } r < \sigma$ \cr
                 0, & ${\rm if\ } r \geq \sigma$ , \cr }
\label{eqn:step}
\end{equation}
\begin{equation}
U_{\rm gm}(r) = -kT \ln (1 - e^{-r^2/\sigma^2}) .
\label{eqn:sh}
\end{equation}
\end{mathletters}
The limit $\beta = 1/kT
\rightarrow \infty$ for the step potential corresponds to the
Widom-Rowlinson model. For the Gaussian molecule potential, the
temperature $T$ plays no role because the Boltzmann factor,
$e^{-\beta U_{\rm gm}(r)}= 1-e^{-r^{2}/\sigma^2}$ is, by design,
independent of $T$.

In general, each component of the $q$-component fluid may have a
distinct fugacity; however symmetry considerations dictate that a
demixing transition occurs with all fugacities
equal, and hence we
set all the fugacities equal to a single value, $z$. For
sufficiently small
$z$, the $q$ species are mixed, while for large $z$, there are $q$
distinct phases because different species repel one another, with
each phase predominately composed of one species. If $q$ is not
too large, there is expected to be a single demixing transition
separating these regimes that is in the same universality class as
the $q$-state Potts model.

For very large $q$, the correspondence between Potts models and
Widom-Rowlinson models must break down. Although Potts models have
a single ordering transition, Widom-Rowlinson models can be
presumed to have an intermediate crystalline phase for $d\geq3$
and large $q$. To understand this phase, consider the limits $z
\ll 1$ and $q \gg 1$, with the product $\lambda = qz$ order unity. 
Then non-overlapping particles appear with an effective fugacity
of $\lambda$. However when two particles overlap, the cost is an
additional factor of $1/q$ because the overlapping particles must
be of the same species. Hence the limiting model is precisely the
hard sphere gas which we presume has a crystalline phase in
$d\geq3$\cite{frisch}. It is therefore reasonable to assume that
such a phase occurs in the Widom-Rowlinson models for large $q$
and $z$ of order $q^{-1}$. Needless to say, for fixed $q$, when
the fugacity is sufficiently large, the model will demix, and thus
the crystalline phase is an intermediate phase. We also expect an
intermediate crystalline phase for large $q$ soft-core
\SH\ models based on a mapping to a single component fluid with a
repulsive soft-core potential. For example, the Gaussian core
model\cite{StWe} is known to crystallize in $d=3$. Although
intermediate phases do not occur for the usual Potts models, they are
not a consequence of the continuum; indeed such phases are known to
occur on the lattice for the site-dilute (annealed) Potts
models\cite{CKS} as well as for the lattice version of the
Widom-Rowlinson model\cite{RL}.

\section{Cluster Algorithm}
\label{sec:methods}

The algorithms used here are, broadly speaking, examples
of cluster algorithms of the type first introduced by Swendsen
and Wang\cite{SwWa,Newbook}. Cluster algorithms have been found
to be much more efficient than local algorithms such as the
Metropolis algorithm for simulating spin systems and lattice
gases near critical points. Cluster algorithms would be very
useful for off-lattice systems, but no general cluster method has
yet been developed; indeed, the {\em only} off-lattice models for
which highly efficient cluster methods are known are models of the
Stillinger-Helfand and Widom-Rowlinson type. The distinguishing
features of this class of models are that particles of the same
species have no self-interaction and that there is a purely
repulsive interaction between particles of different species. In
this case, graphical representations and clusters algorithms are
available\cite{CM,HLM} and have been implemented for the
Widom-Rowlinson model\cite{JGMC}.

Cluster algorithms for spin systems work by identifying clusters of
spins and then randomly flipping these clusters. Cluster are defined
by placing bonds between nearest neighbor aligned spins with a
probability that depends on the temperature. For fluid systems,
bonds are placed between particles of the same type with a
probability that depends on the temperature and the interaction
potential. Instead of flipping spins, clusters of particles are
removed from the system and then new particles are added via a
nonuniform Poisson process that depends on the fugacity and the
potential due to the remaining particles.

Cluster algorithms are typically used with fixed values of the
external parameters such as temperature or fugacity. However, when
the location of the phase transition is not known, much
computational effort in studying the transition is spent in
locating the transition. To avoid this problem, invaded cluster
methods can be used\cite{MaCh95,MaCh96} which
automatically adjust a thermodynamic parameter (for example,
temperature or fugacity) to its value at the phase
transition. This adjustment is accomplished by using the fact
(proved for the
$q=2$ case\cite{CM}) that the clusters just percolate at the
transition. In invaded cluster algorithms, clusters are grown
until a signature of percolation is observed. The value
of the thermodynamic parameter at the transition is an output
of the simulation obtained from the fraction of successful attempts
to add particles or bonds to the system. The invaded cluster
algorithm also may be used to distinguish first-order from
continuous transitions as discussed in Ref.\cite{MaCh96} for Potts
models. This method for distinguishing the order of the transition
is discussed below and will be used in Section~\ref{sec:stepq}.

We first describe the cluster algorithm discussed in Section~3.5
of Ref.\cite{CM} for
\SH\ models and then discuss how it can be modified to be an
invaded cluster algorithm. We assume that we have a configuration
consisting of particle positions and a set of bonds connecting
some of the particles and describe how to obtain the next
configuration:

\begin{enumerate}
\item Identify all clusters of particles defined by the bonds. A
particle with no bonds is considered to be a singleton cluster.
For each cluster, independently and with probability
$1/q$, label it a {\em black} cluster and with probability
$1-1/q$ label it {\em white}.

\item Remove all particles in black clusters. The remaining white
particles are at a set of positions $W$.

\item Replenish the black particles via a Poisson process with
local intensity $y(x)$ given by
\begin{mathletters}
\begin{equation}
\label{eq:intensity}
y(x) = z e^{-\beta V(x)},
\end{equation}
\begin{equation}
V(x)=\sum_{y \in W} U(|x-y|) .
\end{equation}
\end{mathletters}
where $z$ is the fugacity and $U(r)$ the potential.
\item For each pair
of black particles,
place a new {\em bond} between them with probability $p(r)$ given by
\begin{equation}
p(r) = 1 - e^{-\beta U(r)},
\end{equation}
where $r$ is the separation between the particles. Note that $p(r)$
is minus the Mayer $f$-function for the potential.

\item Eliminate the white and black labels for the clusters.

\end{enumerate}
This procedure comprises one Monte Carlo step.

Given a configuration of particle positions and bonds without
species labels, it is possible to obtain a full multicomponent
configuration where each particle has a species label. This
assignment is accomplished by identifying clusters and then
randomly and independently assigning one of the
$q$ species labels to each cluster. The species label of each
particle is taken to be the species label of its cluster. This
labeling of particles is only possible if $q$ is a positive
integer. However, the algorithm makes sense for all $q
\geq 1$, in analogy to the relation between Potts models, which
are defined for positive integer
$q$, and random cluster models which interpolate between them and
are defined for all $q>1$.

It is instructive to consider the nature of the cluster
configurations generated by the algorithm as a function of the
fugacity for a fixed temperature. Suppose that the fugacity and,
hence, the density is very small. Then $p(r)$ is typically small
because the particles are far apart, and most clusters are
singletons. In step 2, a fraction
$1/q$ of the particles is removed. In step 3, particles are
replenished as a nearly ideal gas because the exponential factor
in Eq.~(\ref{eq:intensity}) is a small perturbation except in the
vicinity of the remaining particles. The end result is a nearly
ideal multi-component gas. In the limit of large
fugacity and density, we expect a phase in which a
single species is predominant with a small admixture of the other
species. The bonds connecting particles of the dominant species
are sufficiently dense that almost all members of this species
are in a single large cluster. The minority species are almost all
in widely scattered singleton clusters. When the majority species is
white, as occurs in about $1/q$ of the Monte Carlo steps, the large
cluster is removed and then replaced as a nearly ideal gas in a
slightly perturbed background potential generated by the minority
species. An important feature of this picture is that the
clusters do not percolate at small fugacity and do percolate at
large fugacity. At some intermediate value of the fugacity,
there must be a percolation transition. As discussed in
Refs.\cite{CM,Kl,ChChKo},
the percolation transition of
the clusters coincides with the demixing transition of the fluid.

The coincidence of the percolation transition and the demixing
transition justifies an invaded cluster version of the above cluster
algorithm. The invaded cluster algorithm is very similar to the
fixed $z$ cluster algorithm described above except that steps 3 and
4 are modified as follows. Instead of putting down new black
particles as a Poisson process at a fixed intensity, black
particles are added to the system one at a time according to the
potential
$V(x)$ (see below). After each black particle is added, bonds
between the new particle and all previously placed black particles
are put down with probability $p(r)= 1 - \exp(-\beta U(r))$. The
black clusters defined by these bonds are monitored after each
particle is added, and the process of adding particles is stopped
when a {\em stopping condition} is satisfied. For simulating the
phase transition, the stopping condition is that one cluster {\em
spans} the system. For periodic boundary conditions, spanning is
taken to mean that a cluster wraps around the system in at least
one of the
$d$ directions. The spanning
condition insures that the algorithm simulates the phase
transition\cite{MaCh96}.

In practice, a particle is added to the system according to the
potential $V(x)$ by the following procedure. A particle is
tentatively placed at a random position $x$. A random number $r$
is chosen in the interval $[0,1)$, and the particle placement is
accepted if
\begin{equation}
r < e^{-\beta V(x)};
\end{equation}
otherwise the particle is rejected and another
attempt is made to place a particle. Let
$\z=\langle N_{\rm tot}/L^d \rangle$, where
$N_{\rm tot}$ is the total number of attempted particle placements
in a Monte Carlo step, including both accepted and rejected
placements,
$L^d$ is the system volume, and the brackets $\langle
\ldots \rangle$ indicate an average over the simulation. Because
the intensity
$y(x)$, defined in Eq.~(\ref{eq:intensity}), and the Boltzmann
factor
$e^{-\beta V(x)}$ governing particle placements differ by a factor
of the fugacity, we conclude that $\z$ is an
estimator of $z_{c}$, the value of the fugacity at the
transition. Note that if the fluctuations $\sigma_{\z}$ in $N_{\rm
tot}/L^d$ are small, then the invaded cluster algorithm is
essentially identical to the fixed fugacity algorithm operating at
$z= z_c$. This identification justifies the use of the
invaded cluster method. A more complete discussion of the
invaded cluster method and the use of $\z$ as an estimator of a
critical parameter is given in Ref.\cite{MaCh96}.

Whenever the invaded cluster method simulates a system at its
critical point, scaling methods can be used to obtain
critical exponents from the size dependence of divergent
thermodynamic quantities such as the compressibility or the
susceptibility. To study the latter, we consider the quantity
\begin{equation}
\label{eq:chidef}
\chi \equiv \frac1{L^d}\langle \sum_i s_i^2 \rangle
\end{equation}
where $s_i$ is the number of particles in the
$i$th cluster. We now show that $\chi$ is related to the usual
susceptibility. Consider, for
simplicity, the discretized version of the \SH\ model on a lattice
of linear dimension $L$ with spacing
$\epsilon$ so that the total number of sites is
$[L/\epsilon]^d$. The demixing order parameter at site $x$ is
given by
$\delta \rho_1(x) \equiv n_1(x) - n(x)/q$, where $n_1(x)=1$
if there is a particle of type 1 at site $x$ and $n_1(x)=0$
otherwise; $n(x)$ counts the presence of a particle of {\it
any} type. The relevant susceptibility $\tilde \chi$ is defined by
the second derivative of the pressure with respect to the
(ordering) chemical potential:
\begin{equation}
\label{eq:chidef}
\tilde \chi = \frac1{L^d}\sum_{x,y}
\langle \delta\rho_1(x) \delta\rho_1(y) \rangle.
\end{equation}
(The reason that $\epsilon$ does not enter explicitly into
Eq.~(\ref{eq:chidef}) is that the derivatives are with respect to
the log of the activity and it is the activity that is scaled by
$\epsilon$.) For a given particle and bond
configuration, averaging over assignments of species labels, it is
clear that
$\delta\rho_1(x) \delta\rho_1(y)$ vanishes unless the sites $x$ and
$y$ are both occupied and in the same cluster, in which case the
result is $q^{-2}(q-1)$. Thus, for a fixed
particle-bond configuration, we obtain the number of particles in
the cluster at
$x$ if we sum over
$y$. Summing over $x$ yields the sum of the squares of the cluster
sizes so that
$\tilde
\chi=q^{-2}(q-1)\chi$, and hence we conclude that
$\chi$ is related to the usual susceptibility. Finally, finite
size scaling predicts that
\begin{equation}
\label{eq:chifss}
\chi \sim L^{\gamma/\nu},
\end{equation}
so that the scaling of $\chi$ with system size can be used to
extract the magnetic exponent $\gamma$.

Cluster methods also may be used to distinguish first-order from
continuous transitions. For this purpose, a fixed density stopping
rule is used. Black particles are added to the system until the
density
$\rho$ reaches a fixed value and then $\z$ is measured. In this
way the canonical ensemble is simulated rather than the grand
canonical ensemble. This procedure is done for a range of densities
near the transition. If the
transition is continuous, the fugacity is a strictly increasing
function of
$\rho$. However, if the transition is first-order, then the fugacity
does not increase monotonically with increasing $\rho$ in the
coexistence region. Why does the nature of the $\z$ versus
$\rho$ curve signify whether a transition is continuous or
first-order? Suppose that the demixing transition of a
$q$-component system is first-order. At the transition, there
is coexistence of
$q+1$ phases; $q$ demixed phases and one mixed phase. Because the
repulsive interaction is reduced for the demixed phases, these
phases have a higher density than the mixed phase. Thus, in the
thermodynamic limit, there is a range of $\rho$ for which the
fugacity is constant. Let $\rho_1$ be the density of the mixed
phase and $\rho_2$ the density of the demixed phase. Because
$\ln z =
\partial s/\partial
\rho$, where
$s$ is the entropy density, we have that $s$ is a
linear function of
$\rho$ in the coexistence region. More specifically, $s(\rho)$ is a
linear combination of $s(\rho_1)$ and $s(\rho_2)$, the entropy
densities of the mixed and demixed phases. The linearity of
$s(\rho)$ applies in the thermodynamic limit. However, for a finite
system, the entropy density is not linear in the coexistence
region.
Consider a system with linear
dimension
$L$ and periodic boundary conditions at density
$\rho$. This system also can be viewed as an infinite system with
periodic constraints on the particles. Let
$s(\rho,L)$ be the entropy density of this periodically constrained
system. Now suppose the constraints are removed and the system comes to
equilibrium. If
$\rho_1 \leq \rho
\leq
\rho_2$, demixing will occur spontaneously so that $s(\rho,L) \geq
s(\rho)$ with the equality holding only at the endpoints of the
coexistence range. Because $\ln z = \partial s/\partial \rho$, we
must have that $z$ is non-monotone in the coexistence region. This
approach for distinguishing the order of a transition is very
similar to the microcanonical Monte Carlo method used in
Ref.\cite{GrEcZh}.

\section{Results}

In Section~\ref{sec:gaussian} we present results for the 2D and 3D
\SH\ Gaussian molecule models. The two-component step potential
model is discussed in Section~\ref{sec:step2} and the $q$-component
step potential is discussed in Section~\ref{sec:stepq}.

\subsection{Gaussian molecule model in two and three dimensions}
\label{sec:gaussian}

We simulated the Gaussian molecule model (with the potential
$U_{\rm gm}$ defined in Eq.~(\ref{eqn:sh})) using the invaded
cluster method and the spanning rule described in
Section~\ref{sec:methods} for a range of linear
dimensions $L$ up to 140 in $d = 2$ and 40 in $d=3$. We choose
units such that distances are measured in units of
$\sigma$. We collected statistics for the number of particles in
the spanning cluster
$M$, the critical density
$\rho$, the susceptibility $\chi$, the estimator of the critical
fugacity $\z$ and its standard deviation $\sigma_{\z}$, and the
normalized autocorrelation function for the spanning cluster size,
$\Gamma_M$.
For each value of $L$ we averaged over $10^{5}$ Monte Carlo steps.
The estimator of the critical density is the average
number of particles (of any species) per unit area (volume) when
the spanning condition is fulfilled. Although
$U_{\rm gm}(r)$ does not go to zero at finite $r$, it
becomes very small for larger $r$ and to speed the calculation, we
set $U_{\rm gm}(r)=0$ for $r\geq 3$.

Tables~\ref{table1} and \ref{table2} show the $L$ dependence of
$M$,
$\rho$,
$\chi$, $\z$, $\sigma_{\z}$, and $\tau_M$ for the 2D and
3D
\SH\ models, respectively. The integrated autocorrelation time
$\tau_M$ is defined by
\begin{equation}
\label{eq:tau}
\tau_M = {1 \over 2} + \sum_{t=1}^\infty
\Gamma_M(t).
\end{equation}
This time is approximately the number of Monte Carlo steps
between statistically independent configurations and enters into
the error estimate for $M$. In practice, $\Gamma_M(t)$ becomes
indistinguishable from the noise for $t \approx 10$ Monte
Carlo steps, and it is necessary to cut off the upper limit of the
sum defining
$\tau_M$ when the magnitude of
$\Gamma_M$ becomes comparable to its error.

Note that the fluctuations $\sigma_{\z}$ in $\z$
decrease with increasing $L$ and that $\tau_M$ is
small and hardly increases with $L$. These results
demonstrate the validity and efficiency of the invaded
cluster algorithm. The decrease in $\sigma_{\z}$ shows that as
$L$ increases, the invaded cluster becomes essentially equivalent
to a fixed parameter cluster algorithm for which detailed balance
can be proven.

The error estimates for all quantities in Tables~\ref{table1} and
\ref{table2} except $\tau_M$ were obtained by
computing the standard deviation of the quantity of interest and
dividing by the square root of the number of measurements. This
error estimate does not take into account correlations between
successive Monte Carlo steps. To account for correlations, the
error estimates in the tables for an observable $O$ must be
multiplied by $\sqrt{2
\tau_O}$, where
$\tau_O$ is the integrated autocorrelation time for $O$.
The statistical errors for quantities derived from fits such as
$\rho_c$ and $\gamma/\nu$ include the factor $\sqrt{2
\tau}$ except that $\tau_O$ is replaced by $\tau_M$.

\begin{figure}[h]
\centerline{\epsfxsize=5in \epsfysize=2.8in \epsfbox{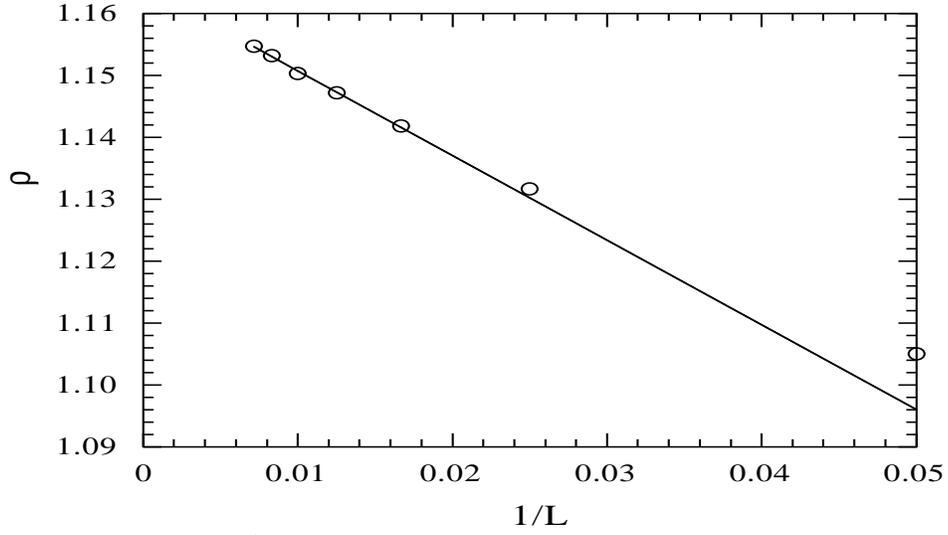}}
\caption{Plot of $\rho$ versus $1/L$ for the 2D \SH\ Gaussian
molecule model. The straight line represents a least squares fit
omitting the value for $L=20$ and $40$. The extrapolated result is
$\rho_c = 1.1644 \pm 0.0004$. }
\label{fig:rho2d}
\end{figure}

Figure~\ref{fig:rho2d} shows the results for
$\rho(L)$ versus $1/L$ for the 2D \SH\ model. The value of $\rho$
in the limit of $L \to \infty$ is extrapolated from the finite size
data by doing a linear least squares fit omitting the values for
$L=20$ and
$40$ yielding the result,
$\rho_c(2)=1.1644 \pm 0.0004$. A similar
extrapolation for the critical fugacity yields
$z_c(2) = 1.3536 \pm 0.0008$. Similarly, extrapolating the result
for the 3D
\SH\ model using the data for all available $L$ yields
$\rho_c(3)=0.440 \pm 0.001$ and $z_c(3)=0.5826 \pm 0.0013$.
Our
error values for these critical parameters are one standard
deviation from the linear least squares fit of
the fugacity or density versus $1/L$; no effort has been made to
estimate systematic errors. All of the fits have acceptable
goodness-of-fit probability values
$Q$.

Our 3D value for the critical density is consistent with the series
result of Lai and Fisher\cite{LaFi},
$\rho_c(3)=0.441 \pm 0.001$ (Eq.~(36) of Ref.\cite{LaFi}) but our
critical fugacity is somewhat larger than their value,
$z_c(3)=0.5785 \pm 0.0002$ (Eq.~(44) of Ref.\cite{LaFi}). Note
that Lai and Fisher report results using a different convention so
that their values of
$\rho_c$ and
$z_c$ must be divided by $\pi^{d/2}$ to compare with our values.

\begin{figure}[h]
\centerline{\epsfxsize=5in \epsfysize=2.8in \epsfbox{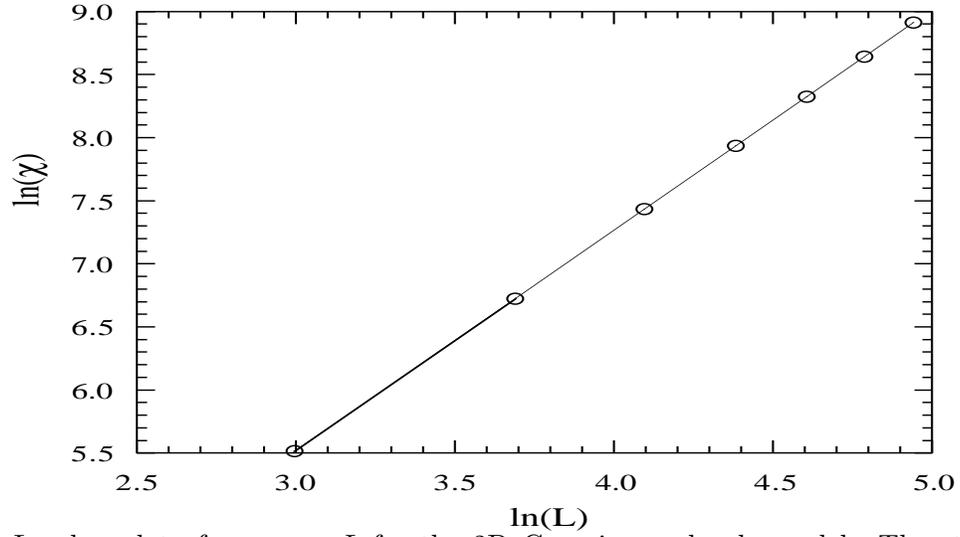} }
\caption{Log-log plot of $\chi$ versus $L$ for the 2D Gaussian
molecule model. The straight line represents the best
fit to the data, omitting $L=20$, with $\gamma/\nu=1.745$.}
\label{fig:chi2}
\end{figure}

The exponent ratio $\gamma/\nu$ can be obtained from the scaling
of the susceptibility $\chi$ with $L$ according to
Eq.~(\ref{eq:chifss}). Figure~\ref{fig:chi2} shows a log-log plot
of $\chi$ versus $L$ for the 2D Gaussian molecule model. A least
squares fit of all the data to a simple power law does not yield an
acceptable goodness of fit value $Q$. If the smallest value of $L$
is omitted, we obtain
$\gamma/\nu = 1.745 \pm 0.001$ with $\chi^2=6.2$,
$Q=0.19$, and ${\rm DF}=4$ (degrees of freedom). The $Q$ value
indicates a reasonable fit to a simple power law, but the fitted
value of
$\gamma/\nu$ is
$5\,\sigma$ from the 2D Ising value of
$\gamma/\nu =7/4$. A reasonable explanation of this result
is that the 2D Gaussian molecule is, indeed, in the 2D Ising
universality class, but that there are relatively slowly varying
corrections to scaling.

A least squares fit to all the data for the susceptibility $\chi$
for the 3D Gaussian molecule model yields
$\gamma/\nu= 1.9626 \pm 0.0044$ with $\chi^2=0.11$, ${\rm DF}=2$,
and $Q=0.95$. The $Q$ value near unity suggests that the data is
well fit to a pure power law. Recent high precision Monte Carlo
studies of the 3D Ising model\cite{BlLuHe} yield
$\gamma/\nu = 1.9630(30)$ which is consistent with our results. Our
results add weight to the hypothesis that the \SH\ model is in the
Ising universality class for both 2D and 3D. The relatively high
precision results for $\gamma/\nu$ from the Gaussian
molecule model suggests that models of the
\SH\ type may be useful for high precision studies of the 3D Ising
universality class. The isotropy of the interaction and absence of
an underlying lattice might make for smaller corrections to scaling
in \SH\ models compared to lattice spin models.

\subsection{Step potential in 2D}
\label{sec:step2}

Table~\ref{table3} summarizes our results for the 2D
\SH\ model with the step potential given by Eq.~(\ref{eqn:step})
and temperature $T=1$ (measured in units of $U_0$). For each value
of $L$ we averaged over $10^{6}$ Monte Carlo steps. The results
are qualitatively similar to the Gaussian molecule model.
Table~\ref{table4} shows the temperature dependence of the measured
quantities for
$L$ fixed at
$L=20$. The values of $\rho$ and $\z$ at low temperatures
should reduce to the Widom-Rowlinson model. In
Ref.\cite{JGMC} we measured the critical parameters of the
Widom-Rowlinson model using the invaded cluster method. For $L=40$
(the smallest size measured) we obtained
$\rho=1.525$ and
$\z=1.720$, values that are close to the values of
$\rho$ and
$\z$ for the two lowest temperatures in Table~\ref{table4}. This
agreement confirms that the step potential is continuously
connected to the hard core potential.

If the $L=20$ data point is omitted, we obtain from a least
squares fit to the data for the susceptibility $\chi$,
$\gamma/\nu = 1.7434 \pm 0.0009$ with $\chi^2=0.52$, $Q=0.81$,
and ${\rm DF}=3$.

\subsection{Dependence of the order of the transition on $q$ and on
impurities}~\label{sec:stepq}
\begin{figure}[h]
\centerline{\epsfxsize=5in \epsfysize=2.8in \epsfbox{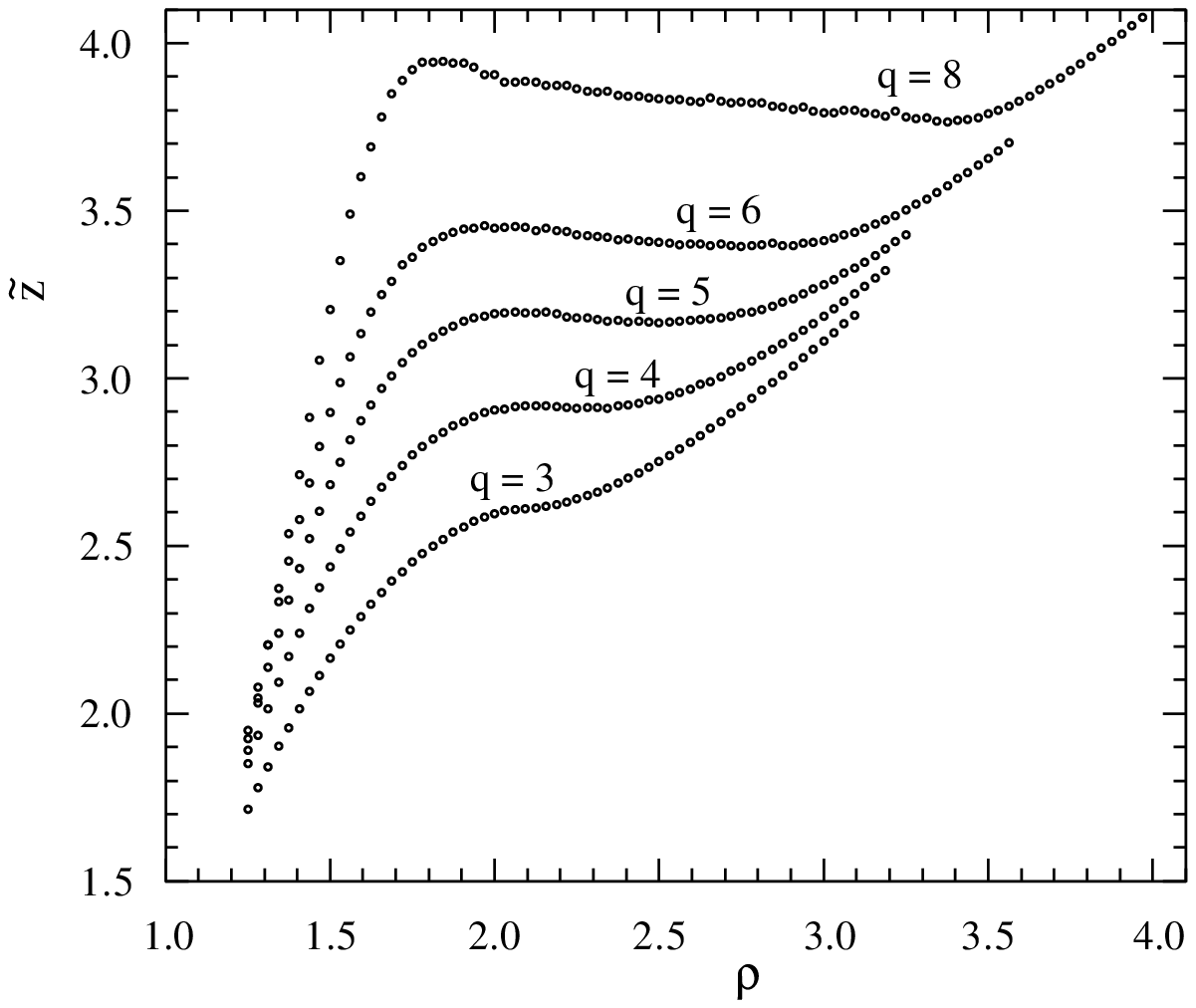} }
\caption{Plot of the fugacity $\z$ versus $\rho$ for the 2D
\SH\ step potential model at $T=1$, $L=40$, for $q=3, 4,
5, 6$, and 8. Each point is averaged over 20,000 MC steps. The
error bars are of the size of the markers.}
\label{fig:fig4}
\end{figure}

The critical properties of the $q$-component \SH\ model are
expected to be closely related to the $q$-state Potts model. One
of the features of the
$q$-state Potts model is that the transition is continuous for
small $q$ and is first-order for $q > q_c(d)$, where $q_c(2)=4$ and
$2 < q_c(3) <3$. We have used the method described in
Section~\ref{sec:methods} to determine the order of the transition
as a function of $q$ for the $q$-component
\SH\ step potential model. Figure~\ref{fig:fig4} shows the
fugacity
$\z$ as a function of $\rho$ for $d=2$
for
$L=40$ and $T=1$. Note that for $q=3$ the curve is clearly
monotonically increasing, which implies a continuous transition.
For
$q
\geq 5$ the curves are clearly non-monotonic, which implies a
first-order transition. For $q=4$ the curve is essentially flat
within the error bars (whose size is approximately that of the
symbols). Although the effective value of $q_c$ is expected to
vary with
$L$, these results are consistent with the hypothesis that
$q_c(2)=4$ for the 2D \SH\ step potential model.

\begin{figure}[h]
\centerline{\epsfxsize=5in \epsfysize=2.8in \epsfbox{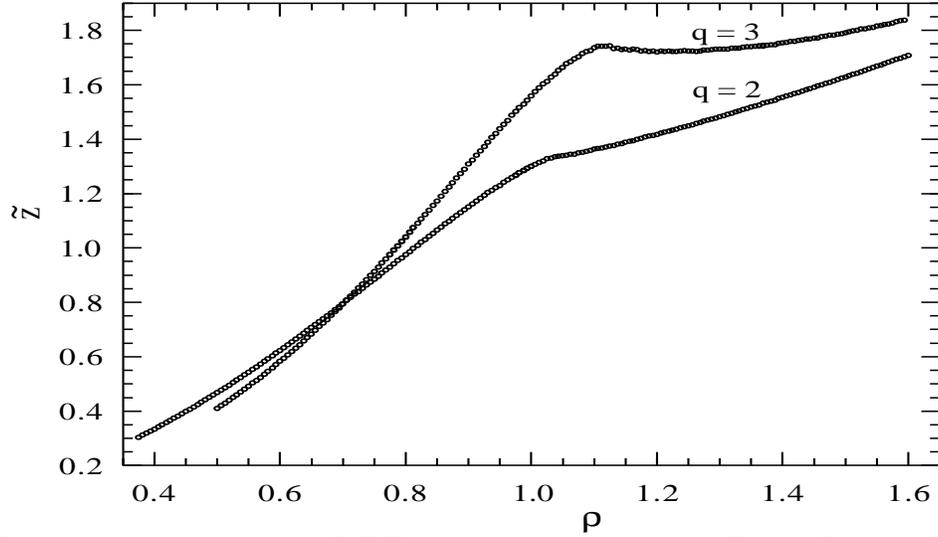} }
\caption{Plot of $\z$ versus $\rho$ for the 3D WR
step potential model at $T=1$, $L=20$, for $q=2$ and 3.
Each point is averaged over 1500 MC steps. The error bars are of
the size of the markers.}
\label{fig:fig5}
\end{figure}

Figure~\ref{fig:fig5} shows $\z$ as a function of $\rho$ for the 3D
\SH\ step potential model for $L=20$ and $T=1$. The $q=2$
curve is clearly monotonically increasing while the $q=3$ is
clearly not, implying that $2 < q_c(3) <3$ as for the 3D
Potts model.

\begin{figure}[h]
\centerline{\epsfysize=3.5in \epsfbox{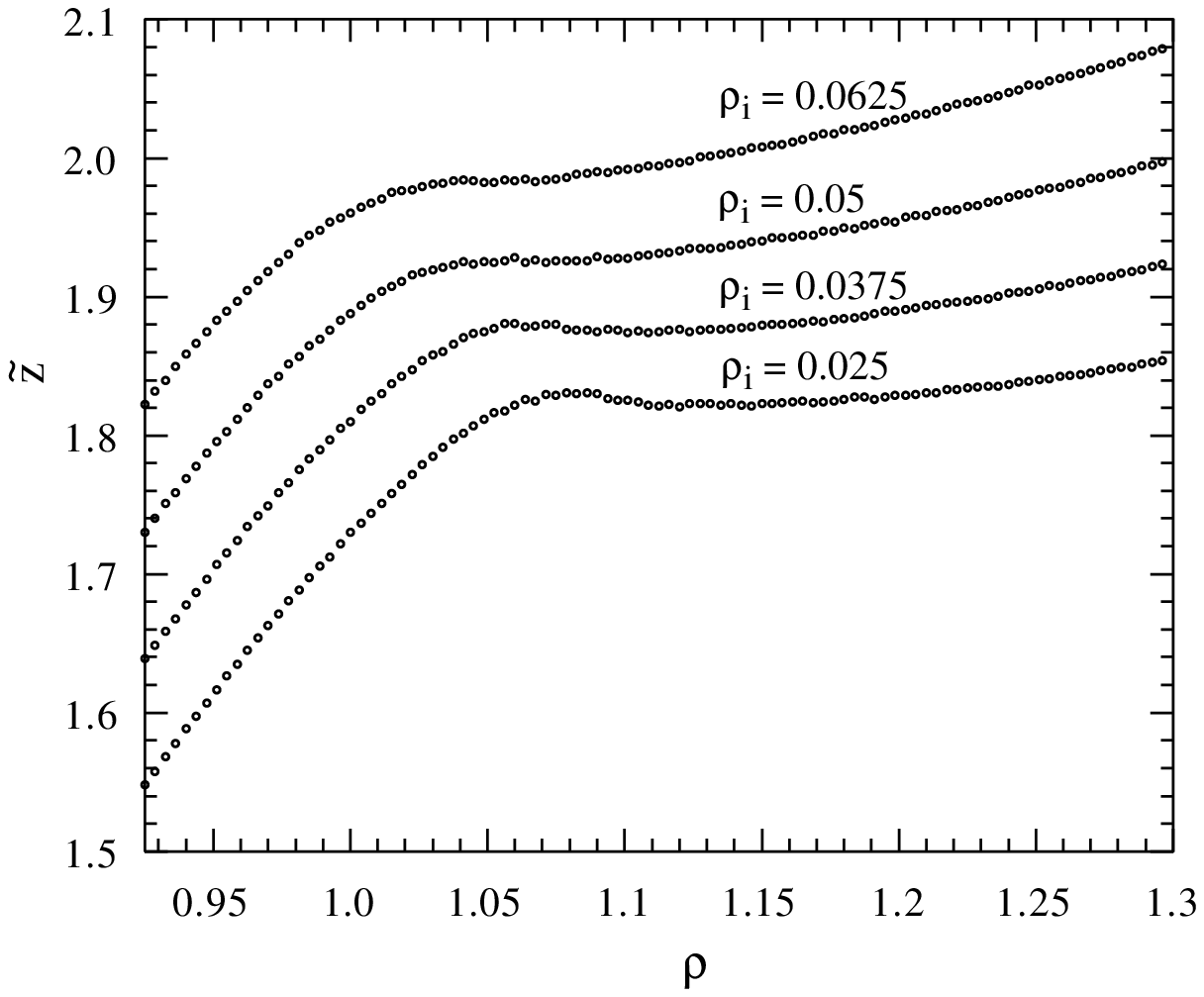} }
\caption{Plot of $\z$ versus $\rho$ for the
$q=3$ \SH\ step potential model in 3D with quenched impurities
at $T=1$ and $L=20$. The 4 traces in the graph correspond to 200,
300, 400 and 500 fixed impurity particles, corresponding to
impurity densities equal to 0.025, 0.0375, 0.05, and 0.0625.}
\label{fig:fig6}
\end{figure}

Finally, we have studied the effect of quenched impurities
on the nature of the transition for the $q=3$ \SH\ step
potential model. The impurities consist of
randomly placed scatterers that interact with all the fluid
particles via the same repulsive step potential that exists between
different components. Figure~\ref{fig:fig6} shows a plot of
$\z$ versus $\rho$ for the
$q=3$ \SH\ step potential model in 3D for four impurity densities
ranging from 0.025 to 0.0625. For each of the 10 impurity
configurations considered for a given density, data from $10^3$
Monte Carlo steps are collected. For the two lowest impurity
concentrations, the
$\z$ versus $\rho$ curve is non-monotonic as is the case for the
pure system, while for the two highest impurity concentrations, the
curve is monotonic indicating a crossover to a continuous
transition. This behavior is in accord with general
arguments\cite{HuBe} that the presence of quenched impurities
should cause a first-order transition to become continuous. It is
not clear from our data whether there is a critical value of the
disorder below which the transition remains continuous or
whether the crossover at finite disorder strength is a finite size
effect and that any strength of disorder is sufficient to make the
transition continuous in the thermodynamic limit.

\section{Discussion and Conclusions}

We have studied the \SH\ model and several generalizations using the
invaded cluster algorithm. Our results for $q$-component \SH\
models with $2 \leq q \leq 8$ are consistent with the hypothesis
that these models are in the same universality class as the
corresponding Potts models. In addition, we have shown that the
addition of quenched disorder causes the demixing transition to
change from first-order to continous for those values of $q$ for
which the pure system transition is first-order. For the case
$q=2$ and
$d=2$, our results for the magnetic exponent are outside the
statistical error bars of the exact Ising value. However, we
believe that this difference is most likely the result of slowly
varying corrections to scaling. It would be useful to consider
larger system to confirm Ising universality. It would also be
interesting to consider larger values of $q$ to explore the
possibility of an intermediate crystalline phase in \SH\ models.

\section{Acknowledgements}
\noindent This work was supported by NSF grants PHY-9801878 (RS),
DMR-9633385 (HG), DMR-9978233 (JM), DMS-9971016 (LC), and NSA grant
MDA904-98-1-0518 (LC). We thank Gregory
Johnson for useful discussions.

\medskip \noindent $^\dag$Present address: Courant Institute of
Mathematical Sciences, 251 Mercer Street, New York, NY 10012.

\newpage

\begin{table}[h]
\caption{The $L$ dependence of the number of particles in the
spanning cluster $M$, the critical density
$\rho$, the susceptibility
$\chi$, the estimator of the critical fugacity $\z$, its standard
deviation $\sigma_{\z}$, and decorrelation time
$\tau_M$ for the 2D Gaussian
molecule potential. The averages are over $10^5$ spanning
clusters. The error estimates were obtained by computing the
standard deviation of the quantity of interest and dividing by the
square root of the number of measurements. The error estimates for
$\tau_M$ are obtained from the variation of $\tau$ with the upper
limit in the summation of Eq.~(\ref{eq:tau}) and hence represent
an estimate of the systematic error. The autocorrelation function
$\Gamma_M(t)$ is distinguishable from the noise for $t \sim 10$
Monte Carlo steps.}
\label{table1}
\begin{tabular}{rcccccc}
$L$  & $M$      & $\rho$     & $\chi$   &     $\z$  &
$\sigma_{\z}$    & $\tau_M$ \\
\tableline
20 &  297.1(1) & 1.105(1) & 245.2(5) & 1.3286(8) & 0.257 &
0.56(1)\\
40 &1095(1) &  1.1317(4)  & 832.5(10) & 1.3469(6) & 0.183 &
0.63(3) \\
60 &2346(2) &  1.1418(3)  & 1695(2)  &  1.3504(5) &  0.149  &
0.68(3) \\
80 &4017(4) &  1.1472(3)  & 2796(2)  & 1.3516(5) & 0.131 &
0.71(5)\\
100&6098(5) &  1.1503(2)  &  4119(3)& 1.3519(4) &  0.117  &
0.75(3)\\
120& 8587(7) &  1.1532(2) & 5670(3) & 1.3524(4) &  0.108  &
0.76(3) \\
140& 11462(10) &  1.1547(2)&7419(5) & 1.3519(4) & 0.0982  &
0.79(2)
\end{tabular}
\end{table}

\begin{table}
\caption{The $L$ dependence of $M$, $\rho$,
$\chi$, $\z$, $\sigma_{\z}$, and $\tau_M$ for the 3D Gaussian
molecule potential. The averages are over $10^5$ spanning
clusters. The error estimates are calculated as discussed in the
caption of Table~\ref{table1}. represents one standard deviation.
The autocorrelation function
$\Gamma_M(t)$ is distinguishable from the noise for $t \sim 60$
Monte Carlo steps.}
\label{table2}
\begin{tabular}{rcccccc}
$L$   & $M$      & $\rho$     & $\chi$   &     $\z$  & $\sigma_{\z}$
& $\tau_M$ \\
\tableline
10 &  171.6(3) & 0.435(2)  &  35.7(2)  & 0.577(2)  &0.103   & 0.52(4)\\
20 &    956(2) & 0.438(1)  & 138.9(4)  & 0.580(1)  &0.0582  & 0.55(3)\\
30 &   2614(3) & 0.438(1)  &   308(1)  & 0.581(1)  &0.0424  & 0.57(2)\\
40 &   5344(6) & 0.439(1)  &   542(2)   & 0.581(1) &0.0341  &0.58(3)
\end{tabular}
\end{table}

\begin{table}
\caption{The $L$ dependence of $M$, $\rho$,
$\chi$, $\z$, $\sigma_{\z}$ and $\tau_M$ for the 2D \SH\ step
potential model at $T=1$. The averages are over $10^6$ spanning
clusters. The
autocorrelation function $\Gamma_M(t)$ is distinguishable from
the noise for $t \sim 10$ Monte Carlo steps.}
\label{table3}
\begin{tabular}{rcccccc}
$L$   & $M$      & $\rho$     & $\chi$     & $\z$   &
$\sigma_{\z}$    & $\tau_M$\\
\tableline
20   &509.5(2)  &1.9249(3)  &720.7(5)   &2.2626(4)  &0.379 &0.585(5)\\
40   &1871.8(6) &1.9633(2)   &2429(2)   &2.2819(3)  &0.270 &0.665(10)\\
60   &4001(2)   &1.9779(3)   &4927(3)   &2.2856(4)  &0.219 &0.702(10)\\
80   &6855(2)   &1.9855(2)   &8131(4)   &2.2865(3)  &0.191 &0.737(10)\\
100  &10411(3)  &1.9902(1)  &12000(6)   &2.2864(2)  &0.169 &0.772(15)\\
120  &14651(4)  &1.9934(1)  &16494(7)   &2.2860(2)  &0.154 &0.783(10)
\end{tabular}
\end{table}

\begin{table}
\caption{The $T$ dependence of $M$, $\rho$,
$\chi$, $\z$, $\sigma_{\z}$, and $\tau_M$ for the 2D \SH\ step
potential model at $L=20$. The averages are over $10^6$ MC steps.
The
autocorrelation function $\Gamma_M(t)$ is distinguishable from
the noise for $t \sim 60$ Monte Carlo steps.
The computational time required for a $L=20$ system of $T=10$ is
approximately a week on a 533 MHz Alpha processor.}
\label{table4}
\begin{tabular}{rcccccc}
$T$   & $M$      & $\rho$     & $\chi$     & $\z$   &
$\sigma_{\z}$    & $\tau_M$\\
\tableline
0.05  &411.1(2)  &1.4905(3)   &470.3(3)   &1.7015(4)  &0.310   &0.58(0.5)\\
0.2   &412.2(2)  &1.4950(2)   &472.4(3)   &1.7072(4)  &0.304   &0.57(0.5)\\
0.5   &439.1(2)  &1.6076(2)   &535.9(3)   &1.8508(4)  &0.326   &0.58(0.5)\\
1     &509.5(2)  &1.9249(2)   &720.6(4)   &2.2626(4)  &0.379   &0.59(1)\\
3     &763.0(3)  &3.3065(4)  &1612.7(8)   &4.1052(6)  &0.598   &0.62(0.5)\\
5     &963.1(7)  &4.6606(4)    &2567(2)   &5.9353(8)  &0.785   &0.64(1)\\
7      &1133(1)  &5.9934(5)    &3551(2)   &7.7433(10) &0.957   &0.65(1)\\
10     &1353(1)  &7.9680(9)    &5059(4)   &10.426(2)  &1.182   &0.66(1)
\end{tabular}
\end{table}

\clearpage

\clearpage

\clearpage

\clearpage

\begin{thebibliography}{1}

\bibitem{StHe} F. H. Stillinger and E. Helfand, {\sl J. Chem.
Phys.} {\bf 41}, 2495 (1964).

\bibitem{StHe2} E. Helfand and F. H. Stillinger, {\sl J. Chem.
Phys.} {\bf 49}, 1232 (1968).

\bibitem{BaMaRo} A. Baram, M. W. Maddox, and J. S. Rowlinson, {\sl
Mol. Phys.} {\bf 76}, 1093 (1992).

\bibitem{LaFi} S.-N. Lai and M. E. Fisher, {\sl Mol. Phys.} {\bf
88}, 1373 (1996).

\bibitem{HuBe} K. Hui and A. N. Berker, {\sl Phys. Rev. Lett.}
{\bf 62}, 2507 (1989); {\bf 63}, 2433 (E) (1989).

\bibitem{AiWe} M. Aizenman and J. Wehr, {\sl Phys. Rev. Lett.}
{\bf 62}, 2503 (1989).

\bibitem{JGMC} G. Johnson, H. Gould, J. Machta, and L. K. Chayes,
{\sl Phys. Rev. Lett.} {\bf 79}, 2612 (1997).

\bibitem{WiRo}
B. Widom and J. S. Rowlinson,
{\sl J. Chem. Phys.} {\bf 52}, 1670 (1970).

\bibitem{frisch}H. L. Frisch and J. K. Percus, Phys. Rev. E {\bf
60}, 2942 (1999).

\bibitem{StWe}
F. H. Stillinger and T. A. Weber,
{\sl J. Chem. Phys.} {\bf 68}, 3837 (1978).

\bibitem{CKS}
L. Chayes, R. Koteck\'y, and S. Shlosman,
{\sl Comm. Math. Phys.} {\bf 171}, 203 (1995).

\bibitem{RL}
L. K. Runels and J. L. Lebowitz,
{\sl J. Math. Phys.} {\bf 15}, 1712 (1974).

\bibitem{SwWa}
R. H. Swendsen and J. S. Wang,
{\sl Phys. Rev. Lett.} {\bf 58}, 86 (1987).

\bibitem{Newbook} M. E. J. Newman and G. T. Barkema, {\sl Monte
Carlo Methods in Statistical Physics}, (Oxford U. Press, Oxford,
1999).

\bibitem{CM}
L. Chayes and J. Machta,
{\sl Physics} A {\bf 254}, 477 (1998).

\bibitem{HLM}
O. Haggstrom, M. N. M. Lieshout, and J. Moller,
{\sl Bernoulli} {\bf 5}, 641 (1999).

\bibitem{MaCh95}
J. Machta, Y. S. Choi, A. Lucke, T. Schweizer, and L. V. Chayes,
{\sl Phys. Rev. Lett.} {\bf 75}, 2792 (1995).

\bibitem{MaCh96}
J. Machta, Y. S. Choi, A. Lucke, T. Schweizer, and L. M. Chayes,
{\sl Phys. Rev.} E {\bf 54}, 1332 (1996).

\bibitem{Kl} W. Klein,
{\sl Phys. Rev. B} {\bf 26}, 2677 (1982).

\bibitem{ChChKo}
J. T. Chayes, L. Chayes, and R. Koteck\'{y},
{\sl Comm. Math. Phys.} {\bf 172}, 551 (1995).

\bibitem{GrEcZh} D. H. E. Gross, A. Ecker, X. Z. Zhang, {\sl
Annalen Phys.} {\bf 5}, 446 (1996).

\bibitem{BlLuHe}
H. W. J. Bl\"{o}te, E. Luijten, and J. R. Heringa,
{\sl J. Phys. A: Math. Gen.} {\bf 28}, 6289 (1995).

\end{thebibliography}
\end{document}